\documentstyle[12pt]{article}
\title{Do Maxicharged particles exist?}
\author{N.V.~Makhaldiani$^{1)}$ and  Z.K.~Silagadze$^{2)}$ \vspace*{3mm} \\
$^{1)}$ \small \em Joint Institute for Nuclear Research, \\
\small \em 141 980, Dubna, Moscow region, Russia \\
\small \em E-mail: \ makhaldiani@main1.jinr.dubna.su \vspace*{2mm} \\
$^{2)}$ \small \em Budker Institute of Nuclear Physics \\
\small \em 630 090, Novosibirsk, Russia }

\date{}

\begin{document}
\large
\maketitle

\begin{abstract}
 The critical charge $Z_c$ is estimated for elementary particles using
a Newton-Wigner position operator inspired model. Particles with $Z \sim
Z_c$ (maxicharged particles), if they exist at all, can have unusual
properties which turn them into illusive objects not easy to detect.
Dirac's magnetic poles have a (magnetic) charge $g\gg Z_c$. This gives one
more argument that it is unexpected that pointlike monopoles to be
found in our world, where $ \alpha^{-1}\simeq 137$.
\end{abstract}

 The aim of this brief note is to raise a question, rather than to give
the answer on it. Why all observed elementary (not composite) particles
have small electric charge $|Z|\leq 1$? May elementary particles with
$|Z|>1$ exist?

 This question can be considered as a one more aspect of the known
charge quantization mystery. Although this quantization can be
understood in the framework of grand unification theories \cite{1} or
even in the Standard model \cite{2}, the most elegant explanation dates
back to Dirac's seminal paper \cite{3} on magnetic monopols. Neither of
these approaches actually exclude the existence of multicharged
particles.

 As small electric charges can more easily escape detection than
big charges, theorists are more willing in introducing the former in
their theories. So in the literature such exotics can be found as
millicharged \cite{4} or minicharged \cite{5} particles. They were
experimentally searched \cite{6}, but not yet found. As for multicharged
particles, only a few (to our knowledge) examples were suggested.
Doubly charged Higgs boson was introduced in \cite{7} and doubly charged
(but composite) lepton in \cite{8}.  Neither of them were found at yet
\cite{9}.

 At least one reason can be imagined which makes big charges
uncomfortable. It is well known \cite{10,11} that, when a nucleus
charge increases, ground state electron energy level in its Coulomb
field lowers and for some critical value of the charge, $Z_c\simeq
170$, plunges into the Dirac's sea of negative energy levels. After
this the vacuum becomes unstable. So $Z_c$ determines an
"electrodynamical upper frontier" for the periodic system of chemical
elements.

 But a finite size of the nucleus, which removes Coulomb field singularity
at the origin, plays an important role in reaching such a conclusion
and in calculation of $Z_c$: The Dirac's equation with bare Coulomb
potential becomes illdefined for $Z>137$. And fundamental elementary
particles (quarks, leptons,...) are believed to be pointlike. So at first
sight the above described notion of critical charge dos not make sense for
them.

 However an arbitrarily precise localization is impossible for a
relativistic particle, as was realized a long time ago \cite{12}.
This means that in relativistic theory an elementary particle no longer
can be considered as a pointlike source for the Coulomb field.

 The meaning of the localization for relativistic particles was
carefully investigated \cite{13,14}. In particular, the most localized
wave-packet for spin zero particle with mass $m$, which does not
contain any admixture of negative frequencies, is given by the
Newton-Wigner wave function \cite{13}
\begin{eqnarray}
\psi (r)\sim(\frac{m}{r})^{5/4}K_{5/4}(mr) \; ,  \label{eq1}
\end{eqnarray} \noindent
where $K_{\nu}(r)$ is a modified Bessel function.

 Unfortunately, $\psi(r)$ in (1), belonging to the continuous spectrum,
is not normalizable and diverges at the origin as $r^{-5/2}$. But it can
not be expected that the one particle picture, which is assumed in (1),
remains valid for distances $r\ll m^{-1}$. Therefore, we may consider the
following simple model for pointlike elementary particle with electric
charge $Ze$,
\begin{eqnarray}
(Ze)^{-1}\rho(r)= \left \{ \matrix {0 \hspace*{33mm}, \; {\rm if} \;
r \leq r_0 \cr Cr^{-5/2}K^2_{5/4}(mr)~ , \; {\rm if} \; r>r_0 } \right . .
\label{eq2} \end{eqnarray} \noindent
 Here $\rho(r)$ stands for charge density at a point $\vec{r}$, and the
constant C is determined from the normalization condition
\begin{eqnarray}
4\pi\int_0^{\infty}\rho(r)r^2dr=Ze \label{eq3} \end{eqnarray}
The cutoff parameter $r_0$ must obey $r_0\ll m^{-1}$. We have somewhat
arbitrarily take $r_0=0.01m^{-1}$. The prescription $\rho=0$
when $r\leq r_0$ is a
reflection of our desire Eq.2 to resemble topological soliton model for
electron \cite{15}. Instead we may take $\rho(r)={\rm const} \equiv
\rho(r_0)$ for
$r\leq r_0$. the results do not change significantly for massive enough
particles and for the lightest particle, still in the realm of our
interest, the difference does not exceed 15\%. Having in mind a
qualitative nature of our argumentation, such subtleties will be left
beyond our care. Note that in \cite{15} $r_0$ coincides with electron
classical radius $(137m)^{-1}$, so giving some justification for our
choice.  If some charge $e_1$ probes the spherically symmetric charge
distribution (2), the potential energy of their interaction is
\begin{eqnarray}
V=-4\pi\alpha\left [\frac{1}{r}\int_0^rx^2\rho(x)dx+
\int_r^{\infty}x\rho(x)dx \right ] \; , \label{eq4} \end{eqnarray}
\noindent
where $\alpha=\frac{Z|ee_1|}{4\pi}$, and opposite sign charges were assumed.

 Now we are inclined to consider Dirac's equation, with the potential
defined from (2$\div$4), for the ground state energy level in the situation
when this level just dived into the negative energy sea, that is E=-1,
in units for which the probe particle mass $m_1=1$. For $m\gg m_1$,
this equation for the radial function G looks like \cite{10}
\begin{eqnarray}
\ddot G-\frac{\dot V}{V}\dot G+\left [V(V+2)+\frac{1}{r}
\frac{\dot V}{V}\right ] G=0 \; ,
\label{eq5} \end{eqnarray} \noindent
where points designate derivatives, for example,$\dot G=\frac{dG}{dr}$.

By substitution $G(r)=\sqrt{V(r)}\psi(r)$, this equation takes the form
which is more convenient for numerical calculations
\begin{eqnarray}
\ddot \psi+\left [V(V+2)+\frac{1}{r}\frac{\dot
V}{V}+\frac{\ddot V}{2V}-\frac{3}{4} \left (\frac{\dot V}{V}\right )^2
\right ] \psi =0 \; . \label{eq6} \end{eqnarray}
For large distances $r\gg m^{-1}$, $K^2_{5/4}(mr)$ in (2) falls as
$e^{-2mr}$. Therefore the second term in (4) can be dropped for such
distances and the first term, because of the normalization condition (3),
gives just the Coulomb potential $V(r)=-\alpha/r$, for which equation (5) is
exactly solvable in terms of the modified Bessel function of complex
index \cite{10}
\begin{eqnarray}
G(r)\sim K_{i\nu}(\sqrt{8\alpha r}) \; , \; \;
\nu=2\sqrt{\alpha^2-1} \; . \label{eq7} \end{eqnarray}
Let us take some $R\gg (2m)^{-1}$. Equation
(5) (in fact(6)) can be numerically solved in the region $0\leq r\leq R $
subject to the boundary conditions G(0)=0, $\dot G(0)\not=0$. Then the
smoothness of the logarithmic derivative at $r=R$ gives an equation which
determines the critical coupling $\alpha_c$:
\begin{eqnarray}
\left . \frac{z \frac{dK_{i\nu}(z)}{dz}}{K_{i\nu}(z)}\right |_{z=\sqrt
{8\alpha r}}
 = \left . \frac{2R\dot G(r)}{G(r)}\right |_{r=R} \; \; .
\label{eq8} \end{eqnarray}
The critical coupling so evaluated shows weak dependence on the mass $m$
and changes from $\alpha_c\simeq 1.03$ for $m=10^4$ to
$\alpha_c\simeq 1.1$ for $m=20$. These numbers correspond to the choice
$R=10m^{-1}$. If we take $R=5m^{-1}$ instead, the modifications don't
exceed a few percent. Roughly modeling particle-antiparticle situation
by setting m=2, we find $\alpha_c \approx 2.5$.

We infer the following main conclusion from the above considerations:
every pointlike electric charge $Ze$, such that $\frac{Z^2e^2}{4\pi}
\approx \frac{Z^2}{137} > 2 \div 3$ destabilizes the vacuum.

The actual value of $\alpha_c$ can be even smaller, if we remember that
field theoretical effects discrease $Z_c$ in the case of nucleus
\cite{16} and some investigations show that chiral phase transition is
expected in strongly coupled QED for $\alpha_c \approx \frac{\pi}{3}$
\cite{17}.

In any case in the following we will treat $\alpha_c \approx 2 \div 3$
as a fair estimate. So $Z_c \approx 15 \div 20$ can be considered as an
"electrodynamical upper frontier" for pointlike elementary particles.

But there is quite a lot space from 1 to $Z_c$. Where are particles
inhabiting this interval?

Particles with $\alpha \approx \frac{Z^2}{137} > 1$ (we will call them
maxicharged particles) are of particular interest, because their
interactions are essentially nonperturbative. For example, an "onium"
from such a particle and antiparticle will decay more willingly into
(n+1) photons than into n photons, because now $Ze>1$. This means that
in fact it decays into an infinite number of soft photons, that is into
a classical field.

Another remarkable property of the maxicharged
particles is that their classical radius $r_0=\frac{\alpha}{m} \;$
($\alpha \approx Z^2/137$) is bigger than their quantum size (Compton
wavelength) $\lambda=\frac{1}{m}$. Because of this property it is not
very easy to produce them in, for example, electron positron
collisions. If $\tau\sim\frac{1}{m}$ is the production time of
maxicharged particle-antiparticle pair and $\tau_0$ their annihilation
time, then \cite{18}
$$\frac{\tau}{\tau_0}\sim\alpha\left(\frac{\lambda}{r_0}\right)^3=
\alpha^{-2} < 1 \; .$$ \noindent
 So the pair is annihilated before they are created \cite{18}! This
suggests that maxicharged particles can be rather illusive objects,
irrespective of their masses.

In fact, the notion of maxicharged
particles was introduced by Schwinger \cite{19}. Below we repeat his
arguments from which more clearly defined and shaped out notion of
maxicharged particles can be deduced.

 Electrodynamics with electric charges e and magnetic charges g reveals
the duality symmetry, which can be viewed as a rotation in the (e,g)
space. However, this symmetry should be spontaneously violated \cite{20},
that is we should have the definite direction for the electric axis in
the (e,g) space. In fact this direction can be guessed from the fact
that the only small charges surround us in our world \cite{19}. First
of all, let us introduce an invariant definition of small
charges \cite{19}: we will say that a particle with electric charge
$e_a$ and magnetic charge $g_a$ belongs to the category of small charges
if
\begin{eqnarray}
\frac{e_a^2+g_a^2}{4\pi}<1 \; .
\label{eq9} \end{eqnarray}
Correspondingly big charges (maxicharged particles in our terminology)
are defined through
\begin{eqnarray}
\frac{e_a^2+g_a^2}{4\pi} \geq 1 \; .
\label{eq10} \end{eqnarray}
 If $a$ and $b$ are an arbitrary pair of small charges, then
\begin{eqnarray}
\bigl(\frac{e_ag_b-e_bg_a}{4\pi}\bigr)^2\leq\frac{e_a^2+g_a^2}{4\pi}~
\frac{e_b^2+g_b^2}{4\pi} < 1 \; .
\label{eq11} \end{eqnarray}
 On the other side, Schwinger's symmetrical quantization condition
reads:
\begin{eqnarray}
\frac{e_ag_b-e_bg_a}{4\pi}=n \; ,
\label{eq12} \end{eqnarray} \noindent
 where $n$ is an integer.

 Now (11) and (12) are compatible only if $n=0$! Therefore, for any pair of
small charges we have \cite{19}
$$\frac{g_a}{e_a}=\frac{g_b}{e_b} \; .$$
 This means that small charges occupy a single line in the (e,g) space,
and it seems from our every day experience that just this line is
chosen as representing electric charge axis after spontaneous breakdown
of the duality symmetry. In other words none of small charges possess
any amount of magnetic charge. Dyons can live only in the wonderland of
maxicharged particles!

Now we turn to more speculative line of reasoning. the most natural
symmetrical solution of Dirac's (nonsymmetrical) quantization condition
$$\frac{eg}{4\pi}=\frac{n}{2}\; , \; \; n-{\rm integer} \; ,$$
\noindent
would be $e=g$. So in such a hypothetical world singly charged particles
will have
$\alpha=\frac{e^2}{4\pi}=0.5$,
and doubly charged particles - $\alpha=2$. Clearly triply charged
particles lay beyond the vacuum stability border, if we adopt the above
cited value for the critical coupling $\alpha_c\simeq 2\div 3$. In fact
even doubly charged particles look suspicious enough. So maybe the observed
absence of multicharged particles is mere reminiscence of the epoch when
there was a full harmony between electrical and magnetic forces?

 Note that the above picture to have any chance to be valid, something
must happen to the \underline {scale} in the duality space, not only to
the orientation of the electric axis, because we know quite well that
$\alpha\simeq(137)^{-1}$ and not 0.5! Can we hope that the present
value of the fine structure constant is associated to the symmetry
breaking between electric and magnetic forces and so can be understood
from purely symmetry considerations? Here we have a tempting
association that just from conformal (or scale) symmetry considerations
Armand Wyler obtained his marvelous formula \cite{21}:
$$\alpha=\frac{9}{16\pi^3}\left(\frac{\pi}{5!}\right)^{1/4}
\simeq\frac{1}{137.03608} \; .$$ \noindent
(For discussions of this formula, see \cite{22}. Some different
"derivations" of this or similar formula can be found in \cite{23},
and for other attempts to calculate the fine structure constant, see
\cite{24}).

Maybe this "number in search of a theory" \cite{25} at last
finds it in electro-magnetic duality and its breaking?

We feel that it is time to finish. Russian folklore says that "one
simpleton can ask so much questions that hundred sages fail to answer".
The only consolation for us is the hope that questions raised in this
essay do not fall into such a category.
\vspace*{5mm}

  The work of one of the authors (N.V.M.) was supported in part by the
U.S. Grant NSF DMS 9418780.

\end{document}